\input harvmac.tex

\line{\hfill PUPT-1638}
\line{\hfill {\tt hep-th/9607113}}
\vskip 1cm

\Title{}{D-Branes and Physical States}

\centerline{$\quad$ {  Sanjaye Ramgoolam and L\'arus Thorlacius }}
\smallskip
\centerline{{\sl Joseph Henry Laboratories}}
\centerline{{\sl Princeton University}}
\centerline{{\sl Princeton, NJ 08544, U.S.A.}}
\centerline{{\tt ramgoola@puhep1.princeton.edu}}
\centerline{{\tt larus@viper.princeton.edu }}

\vskip .3in 

States obtained by projecting boundary states, associated with D-branes, 
to fixed mass-level and momentum generically define non-trivial cohomology 
classes.  For on-shell states the cohomology is the standard one, but when 
the states are off-shell the relevant cohomology is defined using a BRST 
operator with ghost zero modes removed. The zero momentum cohomology falls 
naturally into multiplets of $SO(D-1,1)$. At the massless level, a simple 
set of D-brane configurations generates the full set of zero-momentum 
states of standard ghost number, including the discrete states.  We give a 
general  construction of off-shell cohomology classes, which exhibits a  
non-trivial interaction between left and right movers that is not seen in 
on-shell cohomology.  This includes, at higher mass levels, states obtained 
from typical D-brane boundary states as well as states with more intricate 
ghost dependence.  

\Date{7/96}

\def\npb#1#2#3{{\it Nucl. Phys.} {\bf B#1} (#2) #3}
\def\plb#1#2#3{{\it Phys. Lett.} {\bf B#1} (#2) #3}
\def\prd#1#2#3{{\it Phys. Rev. } {\bf D#1} (#2) #3}
\def\prl#1#2#3{{\it Phys. Rev. Lett.} {\bf #1} (#2) #3}
\def\mpla#1#2#3{{\it Mod. Phys. Lett.} {\bf A#1} (#2) #3}

\def\cmp#1#2#3{{\it Commun. Math. Phys.} {\bf #1} (#2) #3}

\def\bb#1{{\tt hep-th/#1}}

\def\qp{Q^{\prime}}
\def\qpb{ {\bar Q}^{\prime} } 
\def\bB{\bar B} 
\def\qtp{Q_T^{\prime}} 

\lref\pw{J. Polchinski and E. Witten, \npb {460} {1996} {525}, 
\bb{9510169}.}
\lref\polc{J. Polchinski, \prl {75} {1995} {4727}, \bb{9510017}.}
\lref\dlp{J. Dai, R. Leigh, and J. Polchinski, \mpla {4} {1989} {2073}.}
\lref\ck{ C.G. Callan and I.R. Klebanov,  \bb{9511173}.}
\lref\bach{C. Bachas, \plb {374} {1996} {37}, \bb{9511043}.}
\lref\doug{M. Douglas, \bb {9512077}.}
\lref\bsvdt{M. Bershadsky, V. Sadov, and C. Vafa, \npb {463} {1996} 
{420}, hep-th/\-9511222.} 
\lref\gutgr{ M.B. Green and M. Gutperle, \bb { 9604091}.} 
\lref\polcomb{ J. Polchinski, \prd {50} {1994} {6041}, \bb{9407031}.}
\lref\ml{ M. Li, \npb {460} {1996} {460}, \bb {9510161}.}
\lref\Henn{ M. Henneaux, \plb {183} {1987} {59}.} 
\lref\bs{T. Banks and L. Susskind, \bb {9511193}.}
\lref\semiof{ A. Cohen, G. Moore, P. Nelson, and J. Polchinski, \npb 
{281} {1987} {127}.} 
\lref\lzsup{B. Lian and G. Zuckerman, \cmp {125} {1989} {301}.}
\lref\gsw{ M.B. Green, J.H. Schwarz, and E. Witten, {\it Superstring
theory}, Vol. 1, Cambridge University Press, 1987.} 
\lref\deas{S.P. de Alwis and K. Sato, \bb{9601167}.}  
\lref\fof{ J.M. Figueroa O' Farrill and T.Kimura, \cmp {124}{1989} {105}.} 
\lref\bars{I. Bars, \bb {9604200}.}
\lref\clny{ C.G. Callan, C. Lovelace, C.R. Nappi, and S.A. Yost, 
\npb{308} {1988} {221}.} 
\lref\gotho{P. Goddard and C. Thorn, \plb {40} {1972} {235}.}
\lref\fms{D. Friedan, E. Martinec and S. Shenker, \npb {271} {1986} {93}.}  
\lref\polydis{A. Polyakov, {\it Mod. Phys. Lett.} {\bf A6} (1991) 635.}
\lref\witdis{E. Witten, \npb {373} {1992} {187}, \bb{9108004}.}
\lref\disnel{J.Distler and P.Nelson, { \it Nucl. Phys.} {\bf B}  
{\it Proc. Suppl.} {\bf 25A} (1992) 104.}
\lref\zwi{B. Zwiebach, {\it Les Houches 1992 Summer School} 647, 
\bb{9305026}.}
\lref\kt{I.R. Klebanov and L. Thorlacius, \plb {371} {1996} {51}, 
\bb{9510200}.}
\lref\greenscat{M.B. Green, \plb {329} {1994} {435}, \bb{9403040}.}
\lref\cjp{J. Polchinski, S. Chaudhuri, and C.V. Johnson, \bb {960252}.}
\lref\witbd{E. Witten, \npb{460} {1996} {335}, \bb{ 9510135}.}
\lref\oog{H. Ooguri, Y. Oz, and Z. Yin, \bb {9606112}.}
\lref\cardy{J.L. Cardy, \npb {324} {1989} {581}.}
\lref\witM{E. Witten, \npb {443} {1995} {85}, \bb {9503124}.}
\lref\town{P.K. Townsend, \plb {350} {1995} {184}, \bb{9501068}.}
\lref\hull{C.M. Hull, \npb{468} {1996} {113}, \bb{9512181}.} 
\lref\vaf{C. Vafa, \bb{9602022}.} 
\lref\kumat{D. Kutasov and E. Martinec, \bb {9602049}.} 
\lref\sizw{W. Siegel and B. Zwiebach, \npb { 263} {1986} {105}.} 
\lref\lecdis{A. Leclair and J. Distler, \npb {273} {1986} {552}.}    
\lref\comopo{A. Cohen, G. Moore, P. Nelson, and J.Polchinski, \npb {267} 
{1986} {143}.}  
\lref\polcai{J. Polchinski and Y. Cai, \npb {296} {1988} {91}.}
\lref\lzdis{B.H. Lian, G.J.Zuckerman, \cmp {145} {1992} {561}.}
\lref\bezw{ A. Belopolsky, B. Zwiebach, \bb{9511077}.}

\newsec{Introduction}

D-branes \refs{\dlp,\polc} have  played a key role in the 
development of string duality, as they define solitonic states which 
are exchanged with perturbative string states under various duality 
transformations \refs{\polc,\pw}. The study of D-brane boundary states has 
yielded considerable information about D-brane dynamics
\refs{\ml,\ck,\bach,\doug}. In the present paper, we will look for
further insights by studying formal properties of boundary states,
and their relationship to elementary string scattering states.

Boundary states can be usefully viewed as sources
in closed string theory \refs{ \clny ,\polcai }.  
To each worldsheet boundary there corresponds a boundary state and the
leading order effect of including worldsheets with boundaries is to 
add a right hand side to the usual closed string equation of motion:
\eqn\sou{(Q+\bar Q) |\Psi \rangle = | B \rangle \,. }
The nilpotence of $(Q + \bar Q)$ then implies that 
\eqn\ker{ (Q + \bar Q) | B\rangle = 0. }
This equation has been proposed as a general 
criterion that selects admissible boundary states \refs{\clny,\polcomb}.
It suggests that boundary states are physical states of the closed string.
Since $Q+\bar Q$ commutes with $L_0 + \bar L_0$, and with spacetime  
momenta, one can project boundary states to fixed momentum and mass 
level and still have BRST invariant states.  In the following,
we will focus on such projected boundary states, and ask if they 
represent non-trivial cohomology classes.

Consider first the Neumann boundary state of the bosonic 
string in $R^{D-1,1}$
spacetime,
\eqn\neu{|N\rangle = \exp{\left\{ 
-\sum_{n=1}^\infty {1\over n} \alpha_{-n}\cdot\bar\alpha_{-n}
+ c_{-n}\bar b_{-n} + \bar c_{-n}b_{-n} \right\} } 
{1\over 2}(c_0+\bar c_0)|gh\rangle \,,}
where $|gh\rangle$ is annihilated by all positive frequency ghost modes
and all non-negative frequency anti-ghost modes.
More general D-brane boundary states that can be found in the literature
have the same simple ghost structure as $|N\rangle$, but the BRST 
condition \ker\ does not rule out more complicated ghost dependence. 
The specification of the $R^{D-1,1}$ 
 spacetime means that $p_L^{\mu} = p_R^{\mu}$,
while the Neumann boundary conditions require $p_L^{\mu}=-p_R^{\mu}$, 
so the momenta are forced to be zero.    
At higher mass levels, zero spacetime momentum means that 
the states are necessarily off-shell. The usual BRST operator 
does not have off-shell states in its cohomology.
On the other hand there exist interesting amplitudes, for example 
involving closed string exchange between D-branes \refs{\polc,\bs}, 
that are off-shell 
from the closed string point of view\foot{For an early discussion of 
off-shell amplitudes see \comopo\ and references therein.}.  
We are therefore led to look for a definition 
of `off-shell but physical'
states. There is a natural definition in the old covariant formalism,  
given in \gotho . 
We will describe an  adaptation of this to the BRST context, which involves 
a restricted BRST operator, $Q^{\prime}$, with ghost zero modes excised, 
acting on a certain subspace of the Fock space of matter and ghosts.
When this restricted BRST operator is used to define 
off-shell cohomology, both 
on-shell and off-shell components of boundary states belong to non-trivial
cohomology classes. 

We describe the relation between components of a simple class of boundary 
states and the zero-momentum cohomology of the closed string.  Boundary 
states associated with an appropriate set of D-brane configurations 
generate all the cohomology classes at the massless level.  This 
includes `discrete states' which only exist at zero momentum and do not 
appear to play any role in perturbative closed string theory.  The extra 
discrete states at zero momentum combine with those obtained by continuing 
the ordinary perturbative states to form representations of $SO(D-1,1)$.

We then give an algorithm for building closed string off-shell cohomology 
classes at arbitrary mass levels.  By considering zero-momentum states at 
low-lying massive levels, (which are now cohomology classes of 
$Q^{\prime}$ but not of $Q$), we find that simple boundary states do 
not generate the full cohomology.

\newsec{Basic properties of $Q^{\prime}$ cohomology}
\subsec{ Definition of $Q^{\prime} $ cohomology } 
 
In the old covariant formalism there is a natural definition of 
`off-shell but physical' \gotho. Let us recall how that goes, first in the 
open string case. On-shell physical states $|\chi \rangle$ are 
characterized by conditions expressed in terms of the matter Virasoro 
generators $L_{n}^{(m)}$:     
\eqn\osph{\eqalign{ 
& ( L_n^{(m)}  )   |\chi \rangle = 0, {\hbox{ for }} n > 0 \,,\cr    
& ( L_0^{(m)} - 1 ) | \chi \rangle = 0 \,. \cr }}
Off-shell physical states are defined simply by dropping the condition on
$L_0^{m}$. For closed strings we similarly drop the condition 
on $L_0^{(m)} + \bar L_0^{(m)} $.

This definition has a natural extension to the modern covariant
formalism with ghosts present.  Let us make the zero mode dependence
of $Q$ explicit,
\eqn\qexp{ 
Q = c_0 (L_0^{(m)} + L_0^{(gh)} - a ) + b_0 M + Q^{\prime} \,,}
where $M$ is constructed entirely from ghosts,
\eqn\mdef{ M= -2\sum_{n=1 }^{\infty} n c_{-n} c_{n} \,. } 
It follows from the nilpotence of $Q$ that 
\eqn\qpsq{ {Q^{\prime}}^2 = -M (L_0 - a) = - (L_0-a)M }
and that 
\eqn\comu{\eqalign{& [ Q^{\prime}, L_0 ] = 0, \cr   
                   &   [Q^{\prime}, M ] = 0,  \cr  }}
where $L_0 \equiv  L_0^{(m)} + L_0^{(gh)}$.    
The cohomology of the $Q^{\prime}$ operator can be studied,
without loss of generality, on states annihilated by $b_0$. It is 
trivially repeated on states annihilated by $c_0$. 

It follows from \qpsq\ that 
a linear space where ${Q^{\prime}}^2$ is zero may be  selected
by imposing the condition $L_0-a=0$ or by imposing $M=0$.  
If the first  option is chosen, off-shell states are removed from the 
beginning. If the second is chosen, then the cohomology at $L_0-a=0$, 
contains the null states which are $Q^{\prime}$ of states which do not 
satisfy $M=0$. A third option, which we adopt below, is to select 
the appropriate cohomology 
according to whether or not $L_0-a=0$. For $L_0-a \ne 0$, we take 
the cohomology of the  $Q^{\prime}$ operator computed on the space 
satisfying $M=0$. For $L_0-a =  0 $, we take the 
standard $Q$ cohomology (which is closely 
related to $Q^{\prime}$ on $L_0- a= 0 $, 
with no restriction on $M$ \Henn ). In reference \gotho\ it was observed 
that a simple extension of the DDF construction gives off-shell 
physical states satisfying \osph. Here the observation takes the form : 
 
{\bf Proposition 2.1 : } $Q^{\prime}$ cohomology on $M=0$, at $L_0-a\ne 0$ 
contains off-shell continuations of the $DDF$ states. 

These states are annihilated by the positive matter Virasoro modes, 
and are built by acting with  matter operators on the ghost vacuum 
$|gh\rangle$.  For such states, the condition of annihilation by 
$Q^{\prime}$ is precisely  that they should be annihilated by the 
positive matter Virasoro modes. The $DDF$ states contain oscillators
with polarization parallel to a light-like vector, and transverse
oscillators. The analytic continuation of ref. \gotho\
adjusts the momentum along the other  light like direction, without  
changing the coefficients of any oscillators. 
It follows that the  norm of a DDF state 
is unaffected by the continuation. Therefore such states  cannot 
be in $Im (Q^{\prime})$, for if they were, they  would have to be null as 
the following argument shows. Let $|s\rangle = Q^{\prime} |t\rangle $. 
Then the norm is 
\eqn\hermno{ 
 \langle t|( Q^{\prime} )^{\dagger} Q^{\prime} |t\rangle }  
But $Q^{\prime}$ is hermitian, using 
$(c_n)^{\dagger} = c_{-n}$, and $b_{n}^{\dagger} = b_{-n}$, 
so any $\qp$ closed and exact state is null.

We are  interested in the non-chiral version $Q^{\prime}$ cohomology for 
the closed string.  We impose at the outset the conditions 
$L_0-\bar L_0 = 0 $, and $b_0-\bar b_0 = 0$, which are certainly satisfied 
by D-brane boundary states. Defining $Q_T^{\prime}= Q^{\prime} + \bar Q^{\prime}$,
$Q_T=Q+\bar Q$, 
and $L_T=L_0 + \bar L_0$, we have     
\eqn\qprsq{\eqalign{  
  ( Q_T^{\prime} )^2 
& = -M(L_0-a) - \bar M(\bar L_0 -a) \cr
&  =  - {1\over 2} (M+\bar M) (L_0 +\bar L_0 -2a) \cr }} 
In order to have a linear space where ${Q_T^{\prime}}^2$ is zero we need 
to work in the space where $ L_T - 2a= 0$ or the space where 
 $ (M + \bar M) = 0$. Our working  definition of physical states,
will be to consider, 
along the lines of the open string case,  
all states of fixed $L_T - 2a $: if $L_T-2a=0$,
 we use the standard $Q_T$ cohomology, 
whereas if $L_T - 2a \ne 0$, use  
$Q_T^{\prime}$ cohomology at $ (M +  \bar M)  = 0$. 

By arguments similar to the above, this cohomology  contains the 
analytic continuation of the DDF states.
It is easy to see, however, that there are states 
which are  $Q_T$ closed and exact (and hence null) when on-shell, 
but which give non-trivial  $Q_T^{\prime}$ 
cohomology when continued away from mass shell. 
For example two representatives of the cohomology class of the 
dilaton are 
\eqn\dil{\eqalign{  |s^{(1)}\rangle &= \sum_{i=1}^{D-2} 
 \alpha_{-1}^{i} \bar \alpha_{-1i} |gh;p^0=p,p^{D-1}=p \rangle \cr 
|s^{(2)} \rangle &= \sum_{i=0}^{D-1} (\alpha_{-1}^{i} \bar \alpha_{-1i}  
+ c_{-1}\bar b_{-1} + \bar c_{-1} b_{-1}) |gh;p^0=p,p^{D-1}=p \rangle, 
\cr}}
where $|gh;p^0=p ,p^{D-1}=p \rangle$ is a state carrying equal 
non-zero momenta $p^0$ and  $p^{D-1}$, and 
zero momentum in the remaining directions. The difference 
$|s^{(1)}\rangle - |s^{(2)} \rangle $ is $Q_T$ exact. 
However when the momenta 
are continued to $ |p^{0}| \ne |p^{D-1}|$ this state is 
still $\qp_T$ closed, and null, but not $\qp_T$ exact. 
Note that a state which is $\qp_T$ closed and exact is automatically null, a
fact explained by equation \hermno,  
but a state which is $\qp_T$ closed and null is not necessarily exact. 
The fact that it is not exact follows simply from an $sl(2)$ 
structure in the problem (see section 2.6).  
Such null cohomology classes  occur in standard $Q_T$ 
cohomology at zero momentum, but can occur at continuous momenta 
in off-shell cohomology.  They appear in the space spanned by the 
projections of  boundary states. Since they have zero norm, 
they do not contribute to closed string exchange amplitudes, but  
standard BRST decoupling arguments do not prevent them from  
contributing to  higher order amplitudes.  

\subsec{ Boundary states and 
 cohomology} 
The projection of a boundary state onto a given spacetime momentum and
mass-level will generically define an off-shell state, {\it i.e.} one
for which $L_T-2a= l\ne  0 $.  Such a state can be written 
\eqn\sdef{ |s\rangle = A(c_0+\bar c_0) |gh\rangle \,, }
where $A$ represents a string of matter and ghost creation operators,
but containing no ghost zero modes.
The following standard argument shows that a state of this form does
not belong to a non-trivial $Q_T$ cohomology class. 
 
We have $ \{  Q_T , b_0 + \bar b_0 \} =  L_T -2a $.
Consider the state
\eqn\teq{\eqalign{ |t\rangle 
= & {1\over {(L_T-2a)}} (b_0 + \bar b_0 ) A (c_0 + \bar c_0 )|gh\rangle \cr
= & {2 \over l } A |gh\rangle \,. \cr }}
It has the property that $ Q_T |t \rangle  = |s \rangle $.
This proves that $|s\rangle $ is not only $Q_T$-closed, but also $Q_T$-exact. 
It is also easy to see that $(b_0+\bar b_0) |s\rangle $ is 
not in $Q_T$ cohomology. Using $ \{ Q_T, b_0 + \bar b_0 \} = L_T- 2a $, 
we find : 
\eqn\ncl{ Q_T (b_0 + \bar b_0)  |s \rangle =  l |s \rangle \ne 0. }
Thus $ (b_0+ \bar b_0) |s \rangle $ is not in  the cohomology of $Q_T$.

Off-shell components of boundary states do belong to non-trivial 
cohomology classes with  respect to the restricted
BRST operator $Q^{\prime}_T$.  
Since off-shell $Q^{\prime}_T$ cohomology is only found in $Ker(M+\bar M)$,
we first check that $M+\bar M$ annihilates known boundary states. 
For this we only need to consider the ghost part of 
any given boundary state, which is identical to the ghost part 
of the Neumann state $|N\rangle$, for generic open string backgrounds ( 
including 
D-branes).  Applying $M+\bar M$ to the ghost part of $|N\rangle$ gives 
\eqn\mmbar{\eqalign{  
  &  (M+\bar M)  \exp \bigl\{
\sum_{n=1}^{\infty} (  c_{-n} \bar b_{-n} + \bar c_{-n} b_{-n} ) \bigr\}
{1\over 2}(c_0 +  \bar c_0) |gh \rangle \cr
&= \sum_{m=1}^{\infty}  m (c_{-m} c_{m} +  \bar c_{-m} \bar c_{m}) 
\exp \bigl\{ \sum_{n=1}^{\infty} (c_{-n} \bar b_{-n} + \bar c_{-n} b_{-n} \bigr\}
(c_0 +  \bar c_0) |gh \rangle\cr
&= \sum_{m=1}^{\infty}  m (c_{-m} \bar c_{-m} +  \bar c_{-m}  c_{-m})
  \exp \bigl\{\sum_{n=1}^{\infty} (c_{-n}\bar b_{-n}+\bar c_{-n} b_{-n}) \bigr\}
     (c_0 +  \bar c_0) |gh \rangle \cr
&=\, 0 \,. \cr }}
Since $Q_T^{\prime}$ commutes with the ghost zero modes, 
it is clear that $|s\rangle$ is also $Q_T^{\prime}$ closed. 

The following shows that $|t\rangle$ in \teq\ is $Q_T^{\prime}$ closed:
\eqn\clos{\eqalign{ 
Q_T^{\prime} |t\rangle 
& = \bigl( Q_T - 1/2 ( c_0+ \bar c_0) (L_T - 2a ) \bigr )  |t\rangle \cr
& =  |s\rangle  - {1\over 2}(c_0+\bar c_0)(b_0+\bar b_0)|s\rangle \cr
& = 0 \,. \cr }}
In the first line we used the fact that $b_0M $, $\bar b_0\bar M$, and 
$L_0 - \bar L_0$ all annihilate $|t\rangle$.

Now that we have seen that the components of a boundary state are annihilated 
by $Q_T^{\prime}$, it remains to prove that they are not in $Im (Q_T^{\prime})$. 
Suppose that a state is $Q_T^{\prime}$ closed and exact.  Then 
it would have zero norm by \hermno.  
All we have to do, to prove that the projections of the boundary state are 
not $Q_T^{\prime}$ exact, 
is to check that they don't have zero norm.
As an example 
consider a static D-brane in flat spacetime in the bosonic string 
theory.  The norm of the state at level $l$ can be read off from the formula
\eqn\norms{ 
 \langle B| q^{L_0 + \bar L_0} |B\rangle
  = \prod_{n=0}^{\infty} {1\over (1-q^{2n})^{D-2}} \,, } 
which exploits the factorized form of the D-brane boundary state 
as a product of exponentials.  The relevant norm, according to the above 
formula, is manifestly positive.  A similar expression applies for D-brane 
boundary states of the NS-NS sector of the 
superstring.  This guarantees that at each level we have at least one 
cohomology class. In section 2.6 we will give an independent 
argument, which  does not rely on norms, and shows that in 
the off-shell case any $Q_T^{\prime}$ closed 
state of ghost number zero is in cohomology. 

\subsec{Examples at massless level}
There are physically interesting boundary states 
which illustrate, already at the massless level, 
the fact that the components 
of boundary states do not always give 
$Q_T$ cohomology classes, but give $Q_T^{\prime}$ cohomology.  
Consider a boundary state, which acts as a source of closed string 
states winding around a compact direction $X^i$. A non-vanishing value 
of the momentum $p_L^{i}= -p_R^{i}$ defines, in general,
 an off-shell state\foot{An off-shell continuation of this type 
was found useful in \clny\ for regularization purposes.} 
but is compatible with Neumann boundary condition. Such off-shell
winding states are not in the cohomology of $Q_T$ but are in that 
of $Q_T^{\prime}$. 
  
\subsec{Remark  on decoupling of longitudinal states }
For on-shell $S$-matrix calculations a simple formal argument 
can be given for tree level  decoupling of longitudinal states
\fms. We write the longitudinal state as $Q_T|t\rangle$
and then  commute $Q_T$  through the remaining 
operators  until it annihilates the vacuum. In fact 
the on-shell longitudinal states can be written 
as $\qp_T$-exact states, since 
\eqn\extnes{\eqalign{Q_T |t\rangle &= \bigl( Q_T^{\prime} + 
{1\over 2}(b_0+\bar b_0)(M+\bar M) + 
{1\over 2} (c_0 + \bar c_0) (L_T-2a) \bigr) |t\rangle \cr 
&=Q_T^{\prime} |t\rangle.\cr }}
In the first line we have used the level matching condition
$(L_0- \bar L_0 )|t\rangle= 0$ and $(b_0 - \bar b_0)|t\rangle = 0$. 
In the second we have used the on-shell condition and 
$(b_0+\bar b_0)|t\rangle =0$.   
The decoupling argument can be used in the presence of boundary operators 
provided  we use $Q_T^{\prime}$ instead of $Q_T$. 
As was emphasized in \deas, the state that enters into the calculation 
of D-brane amplitudes from the closed string point of view is 
$ (b_0+\bar b_0)|B\rangle$, rather than the $|B\rangle$ that appears 
in the equation of motion \sou. The off-shell components 
of $(b_0+\bar b_0)|B\rangle$ are only annihilated by $\qp_T$, and not by 
$Q_T$. 
In applying decoupling arguments one has to be careful 
about boundaries of moduli space, but it is noteworthy 
that $Q_T^{\prime}$, which gives a particularly simple 
definition of off-shell cohomology, also enters in a formal 
BRST discussion of decoupling in the presence of boundary states.   

\subsec{ Enhanced Lorentz symmetry} 
Standard $Q$ cohomology provides representations of Lorentz symmetries. 
Consider the momentum of an on-shell massless particle, 
$p^{0}= p^{D-1}=p$. The cohomology at this momentum is computed by 
considering the action of $Q$ (which is $SO(D-1,1)$ invariant) on a 
space spanned by oscillators (which have well-defined $SO(D-1,1)$ 
transformation properties) acting on the ground state 
$|gh;p^0=p,p^1=p\rangle$. The ground state is left invariant by an 
$SO(D-2)$ subgroup of $SO(D-1,1)$.  Similarly, the momentum of a massive 
particle is left invariant by $SO(D-1)$.  At zero spacetime 
momentum, however, the cohomology  provides
representations of the full Lorentz group $SO(D-1,1)$
as we illustrate  in the following section. If we consider on-shell 
cohomology,  $p^{\mu}=0$ restricts us to the massless level. Given that 
the $\qp$ (and $\qp_T$) operators are $SO(D-1,1)$ invariant, 
the enhanced symmetry will occur at arbitrary mass level in the case 
of $\qp$ (and $\qp_T$) off-shell cohomlogy.  

\subsec{ Vanishing theorems for off-shell cohomology}
Zero-momentum $Q$ cohomology contains states at exotic ghost numbers, 
and one might expect this to  be a generic feature  of  zero-momentum 
cohomology at higher levels. This is not the case, and off-shell $\qp$ 
cohomology has a surprisingly simple structure.

We will first discuss the open string.  It has been shown \sizw\ that the 
ghost system contains an $sl(2)$ algebra (and that only singlets under 
this $sl(2)$ enter the construction of free string field actions). 
Indeed let $N_{gh}$ be the ghost number operator (without the zero modes): 
\eqn\ghst{ N_{gh} = \sum_{n > 0}( c_{-n}b_{n} - b_{-n} c_{n} ), } 
and $N$ be an operator of ghost number $-2$ defined by 
\eqn\opbb{ N = - \sum_{n>0} {1\over {2n}}  b_{-n}b_{n}.  } 
Then we have the $sl(2)$ relations
\eqn\sltwo{\eqalign{
& [  N_{gh} , M ] = 2M, \cr
& [ N_{gh}, N ]   = -2N, \cr 
& [M , N ] =  N_{gh}, \cr
}}
where $M$ and $N$ are respectively raising and lowering
operators of $sl(2)$.
This allows a simple proof of the following vanishing theorem: 

{\bf Proposition 2.2 :   }
$Q^{\prime}$ cohomology groups, at $L_0-a \ne 0$, 
are empty at non-zero ghost number. 

An off-shell state can only be annihilated by $Q^{\prime}$ if it is 
annihilated by $M$. There are no states in $Ker (M)$ at negative ghost 
number. This is simply because there are  no finite dimensional  
representations of $sl(2)$ with a negative highest spin, and  
at fixed mass level (and momentum) there are only finitely 
many states. Any state $|s_{(n)} \rangle$ of ghost number $n > 0$  
annihilated by $M$ generates a finite dimensional representation when we 
act with the lowering operator $N$. Using the $sl(2)$ relations \sltwo, 
we see that such a state is $\qp$ exact: 

\eqn\imageM{\eqalign{ 
|s_{(n)}\rangle &= {N_{gh} \over n} |s_{(n)} \rangle \cr  
                &=     {MN\over {n}} |s_{(n)} \rangle  \cr
                &= \qp \bigl( {-\qp\over {L_0-1}}  {N\over {n}} 
                   |s_{(n)} \rangle  \bigr ).  \cr }}  
This completes the proof of the proposition. 

Similarly for the bosonic closed string, we have

{\bf Proposition 2.3 : }
$Q_T^{\prime}$ cohomology groups, at $L_T - 2a \ne 0$, 
are empty at non-zero ghost number. 

The proof proceeds as  above, with $M$, $N$, and  $N_{gh}$ 
replaced by $M + \bar M$, $N + \bar N$, and 
$N_{gh} + \bar N_{gh}$ respectively.

Another consequence of the above is that any off-shell ghost number zero 
state $|s\rangle$ annihilated by $Q^{\prime}$ is in $Q^{\prime}$
cohomology. Indeed, suppose 
\eqn\ghzero{ \qp |s \rangle = 0 \quad \hbox{ and } \quad 
|s \rangle = \qp |t\rangle \,, }
then $|t\rangle $ has ghost number $-1$ and $(\qp) ^2 |t\rangle = 0 $, 
which implies that $M$ annihilates $|t\rangle $. But we saw that there are 
no states in $Ker M$ at negative ghost number.  The same  argument 
shows that any off-shell ghost number zero closed string state annihilated 
by $\qp_T $ is in cohomology. 

Identical arguments apply for the NS and NS-NS sectors of open and closed 
superstrings respectively, 
the key feature being the $sl(2)$ symmetry of the ghost and superghost 
system (see Appendix A). 

\newsec{ D-branes and cohomology in bosonic string theory } 
We have seen above that boundary states are not only 
BRST invariant but also generically give rise to non-trivial 
cohomology.  In this section we will consider the massless level 
and zero spacetime momentum. The relevant cohomology is then 
the standard  one, and is larger than at generic momentum. 
`Discrete states' associated with such jumps in cohomology 
were first studied in detail in the context of the  non-critical 
$c \le 1$ string \refs{\polydis,\witdis,\lzdis} where there are infinitely 
many of them. It was later  pointed out \disnel\ that this phenomenon also 
occurs in critical string theories. For a recent discussion of zero momentum 
cohomology see \bezw. 

At the massless level the cohomology is spanned by 
 $(D-2)^2$ states at generic 
momentum. They are obtained in a lightfront  like frame 
by acting with $D-2$  left moving transverse 
DDF operators $A_i$ and  $D-2$ right moving operators $\bar A_i$, 
on  a state carrying momentum  in the 
 $X^0$ and $X^{D-1}$ directions, for example. The DDF operators 
$A_{i}$ and $\bar A_{i}$ carry  spacetime 
indices running from $1$ to $D-2$.
 
At zero momentum however there are $D^2 + 1 $ independent states
in the zero momentum cohomology.  They are 
\eqn\spess{\eqalign{ 
 & \alpha_{-1}^{\mu} \bar \alpha_{-1}^{\nu} |gh\rangle ~~ \hbox{for} 
  ~~ 0 \le \mu,\nu \le (D-1), \cr
&  | D_{gh} \rangle =  
( c_{-1} \bar b_{-1} + \bar c_{-1} b_{-1} ) |gh\rangle .  }}
These states include two  scalars, a traceless symmetric tensor of 
$SO(D-1,1)$ of dimension 
$(D+2)(D-1)/2$, and an antisymmetric tensor of dimension $(D-1)(D)/2$. 
Note that the ghost dilaton is $Q_T$-trivial,
 $ | D_{gh} \rangle = Q_T |\chi \rangle$, with 
\eqn\qt{  |\chi\rangle = 
b_{-1}\bar b_{-1} (c_0 - \bar c_0 )|gh\rangle.}
 On the other hand 
 $|\chi\rangle$ is not annihilated by $b_0 - \bar b_0$, so
$|D_{gh}\rangle$   is  not trivial  
 in the semirelative complex ({\it i.e} in 
 the subspace of the matter-ghost 
Fock space annihilated by    $b_0-\bar b_0$),  which 
defines closed string physical 
states \zwi.

We will see that the projections  of  D-brane boundary states 
to the massless level span  the complete  zero-momentum cohomology.
The boundary state   for a static p-brane with Neumann boundary conditions 
along the directions $i=0$ up to  $p$,  and Dirichlet  boundary 
conditions along $i=p+1$ up to  $D-1$ (=25), has the
 following component at the massless level\foot{In this section we 
find it convenient to work  with the boundary states without the 
$(c_0+ \bar c_0)$ factor.}:
\eqn\bdst{ \bigl( -  \sum_{\mu = 0}^{p}  
\alpha_{-1}^{\mu}(\bar\alpha_{-1})_{\mu} 
    +  \sum_{\mu=p+1}^{D-1} \alpha_{-1}^{\mu} (\bar \alpha_{-1})_{\mu} 
    + \bar c_{-1}b_{-1} + c_{-1} \bar b_{-1} \bigr) |gh\rangle .  } 
Along the Dirichlet directions we should sum over all 
momenta satisfying $p_L=p_R$. Non-zero Neumann 
 momenta satisfying $p_L=-p_R$ are compatible with $\qp_T$ 
invariance and arise when some directions are compactified.

Using \bdst\ we can write the following equations at zero momentum, 
\eqn\dilat{\eqalign{ 
 & |D_M \rangle  =  1/2 (| B (-1) \rangle - | B (D-1) \rangle ), \cr
            & |D_{gh} \rangle  = 1/2 ( | B (-1) \rangle +
 | B (D-1) \rangle ), \cr 
}}
where $|B(p)\rangle$ is the projection of the $p$-brane 
boundary state to the massless level, $|D_M\rangle$ is the 
zero-momentum matter dilaton, and $|D_{gh}\rangle$ 
is the zero-momentum ghost 
dilaton. In this sense the boundary states associated to the 
$(-1)$- and $(D-1)$-brane give rise to the two scalars of 
$SO(D-1,1)$.  

Using the $0$-brane and the D-instanton we can similarly write: 
\eqn\alphze{  \alpha_{-1}^{0}{ ( \bar \alpha_{-1} )_0} |gh\rangle 
=  1/2 \bigl( - |B(0) \rangle + |B(-1) \rangle \bigr). }
Let $|B(1;i)\rangle$ be the $1$-brane, with the direction $i$ 
as the spatial Neumann direction.  Then
\eqn\alphk{ \alpha_{-1}^{i} \bar\alpha_{-1i} |gh\rangle = 
{1\over 2} \bigl(  |B(0)\rangle  - |B(1;i) \rangle \bigr) \,. }   
Now rotate this $1$-brane in the $(ij)$ plane by an angle $\theta$ and 
call the corresponding boundary state $| B(1;ij,\theta) \rangle$.  
Then 
\eqn\askl{ ( \alpha_{-1}^{i}\bar\alpha_{-1j} 
+ \alpha^{j}_{-1}\bar\alpha_{-1i}  ) |gh\rangle = {\partial \over 
{\partial \theta }} |B(1;ij,\theta) \rangle \bigg|_{\theta=0}  }
States with the corresponding antisymmetric combination can be obtained 
by considering $1$-branes with a background magnetic field $F_{ij}$.

States of the form $\alpha_{-1}^{i} ( \bar \alpha_{-1})_0 + 
\alpha_{-1}^{0}\bar \alpha_{-1i}$ can be obtained from boundary states 
of $0$-branes boosted along the $i$ direction \ck\ and the corresponding 
antisymmetric combination can be obtained using a D-string aligned along 
the $i$ direction with a background worldvolume electric field. 

The above states exhaust the set of zero-momentum cohomology classes. 
The states obtained by projecting the various boundary states to the 
massless level are not all linearly independent. We have described one  
generating set made of the D-instanton, the zero brane (stationary and 
boosted), D-strings (with and without constant background electromagnetic 
fields), and the $(D-1)$-brane. Note that both the D-instanton and the 
$(D-1)$-brane are present in this generating set.  Although there is a lot 
of freedom in the choice of generating set, we always need these two, as  
they are the only $p$-brane configurations that give rise to 
the two scalars of $SO(D-1,1)$. 
Given the special properties of the D-instanton in D-brane scattering
\refs{\greenscat,\kt}, it is interesting that it is required in order to 
give the complete set of zero-momentum states.

We close this section with some further comments on the relation 
between D-branes and discrete states. Consider $R^{D-1,1}$ spacetime 
where boundary states will carry zero momentum in each Neumann 
direction. For Dirichlet directions there will be a sum over 
all momenta, but since the BRST operator commutes with momentum,
we can restrict to one momentum at a time, and still have 
BRST invariant states. Precisely at zero momentum for all directions, 
we can form linear combinations from all types of D-branes and extra cohomology 
classes may be expected. 

We can understand why the discrete states do not admit continuation 
to $p_L=-p_R$ by considering toroidal compactifications. 
For instance the boosted D-brane with a boost along the $i$ direction
will only exist when  this direction is non-compact. This  explains 
why states $\alpha_{-1}^0 \bar \alpha_{-1i} + \alpha_{-1}^i  
(\bar \alpha_{-1})_0$ do not admit continuations to $p_L^{i} = - p_R^{i} 
\ne 0$.  Similarly the  boundary states associated with a 
$D$-string with worldvolume  electric field, which generate the antisymmetric 
combinations of timelike and space-like oscillators, do not exist 
when the spatial worldvolume direction of the D-string is compact 
and the electric fields are quantized \witbd.

\newsec{ Boundary states and cohomology in Type II superstrings } 

The definitions of off-shell cohomology developed for 
the bosonic string in section 2 admit a simple generalization 
to type II superstrings as outlined in appendix A. 
Here we will describe generating sets of D-branes for zero-momentum 
cohomology at the massless level. 
In doing this, we will be comparing with 
cohomology computed in a fixed picture as  
in \refs{\lzsup,\fof} and references given there.
In terms of operators which act on the $sl(2)$ 
invariant vacuum this amounts to considering  $Q$-invariant 
operators up to $Q$-commutators, in the space of operators 
consisting of $e^{q\phi}$ times polynomials in $\beta,\gamma$
and their derivatives (and the obvious generalization in the non-chiral 
case) together with matter and ghost operators.
Moreover a simple comparison is possible, which 
closely resembles the bosonic discussion, when we compare 
zero-momentum cohomology classes with the combined set of 
boundary states of D-branes of both type IIA and type IIB superstrings. 

Boundary states contain a sum over an infinite number of pictures
$(q_L, q_R)$ with  $q_L + q_R $ fixed (at $-2$ in the NS-NS sector). 
The sum ensures that the D-brane boundary state is annihilated 
by half the supersymmetries, but as far as BRST invariance is 
concerned there is no need to consider the whole sum. 
In the NS-NS sector for example we can work in the 
$(-1,-1)$ picture and the other terms are related by picture 
changing.  Recall that the boundary state which appears in 
\sou\  is proportional to $(c_0+\bar c_0)|gh \rangle$, but in using the 
boundary states to compute amplitudes one works with $(b_0 + \bar b_0)
|B \rangle$.  So, in general, we will be interested in two types of 
cohomology classes, which have representatives of the form 
\eqn\stdgsti{ X |gh;q_L,q_R \rangle, } 
or 
\eqn\stdgstii{X (c_0 + \bar c_0)  |gh;q_L,q_R \rangle,}
where $X$ is a ghost number zero operator, and $|gh ; q_L , q_R \rangle$ 
is a vacuum state satisfying: 
\eqn\vacs{\eqalign{ &  b_{n}, \bar b_{n} |gh ; q_L , q_R \rangle = 0 
           \hbox { for }  n \ge 0, \cr 
                   &  c_n, \bar c_{n} |gh ; q_L , q_R \rangle = 0 
             \hbox { for }  n > 0, \cr 
                   &  \beta_{n}  |gh ; q_L , q_R \rangle = 0  
\hbox { for } n > -q_L - 3/2,  \cr 
                   & \gamma_n   |gh ; q_L , q_R \rangle = 0 
    \hbox { for } n \ge q_L + 3/2, \cr   
            &  \bar \beta_{n}  |gh ; q_L , q_R \rangle = 0  
\hbox { for } n >  -q_R - 3/2,  \cr 
                   & \bar \gamma_n   |gh ; q_L , q_R \rangle = 0 
    \hbox { for } n \ge  q_R + 3/2. \cr  }}  
We will refer to  the  states in \stdgsti\ and \stdgstii\
as states  of ``standard ghost structure''. 

In performing the comparison between 
cohomology and projections of boundary states we may use, in the 
NS-NS case,  either $|B\rangle $ or $ (b_0+\bar b_0 ) |B\rangle$, 
since the cohomology over $(c_0+\bar c_0)|gh\rangle$ is identical 
to that over the $|gh\rangle$ vacuum. In the R-R and R-NS cases 
this is not always true.

In the NS-NS sector the zero momentum cohomology at standard ghost number
closely parallels  that of  the bosonic string.  Working in the $(-1,-1)$ 
picture, it is spanned by 
\eqn\sds{\eqalign{ 
&   \psi_{-1/2}^{\mu} \bar \psi_{-1/2}^{\nu} 
|gh; -1,-1 \rangle  ,\qquad 0 \le \mu,\nu \le 9 \,, \cr 
& |D_{gh}\rangle =     ( \beta_{-1/2}\bar \gamma_{-1/2}-\bar\beta_{-1/2} 
\gamma_{-1/2 } ) |gh, -1, -1 \rangle \,. \cr}}
Note that in the $(-1,-1)$ picture all the positive $\beta$ and $\gamma$
zero modes annihilate the vacuum. As in the bosonic case, the 
dilaton is $Q_T$ of something which is not annihilated by 
$b_0-\bar b_0$, and is physical in the semirelative complex, 
\eqn\qtag{  |D_{gh}\rangle  = 
          (Q + \bar Q) \beta_{-1/2}\bar \beta_{-1/2} 
                  (c_0 - \bar c_0) |gh, -1, -1 \rangle. } 
Again the  states naturally fall into representations of $SO(D-1,1)$ 
($= SO(9,1)$ in this case).
 
The boundary states of \ml\ are linear combinations of states in the 
NS-NS sector and the R-R sector. They are separately BRST invariant. 
D-branes and anti D-branes differ in the relative sign between the 
two sectors. Thus we can isolate states purely in the 
NS-NS or R-R sector by taking linear combinations.  
In the NS-NS sector all zero-momentum physical states are generated 
by D-branes, at rest or with constant boosts, or constant electromagnetic 
field backgrounds. The argument is identical to the one we gave for the 
bosonic string.

In the R-R sector the zero-momentum limit is more subtle. The boundary 
states in \ml , for example, contain states in the $(-1/2,-1/2)$ picture 
which have projections that vanish in the zero-momentum limit. They can, 
however, be related via picture changing to states in the $(-3/2,-1/2)$ 
picture \cjp, and after this a finite zero-momentum limit may be obtained. 
We hope to give a more detailed account of the relations between boundary 
states and cohomology in the R-R and R-NS sectors in future work. 

\newsec{Construction of off-shell closed string cohomology classes} 
\subsec{General construction}
The condition for $\qp_{T}$ to vanish, in the off-shell case where 
$L_T \ne 0$,   is $M+\bar M =  0 $.
There are many states which satisfy this condition without 
satisfying $M=\bar M= 0$. This allows for a more intricate  
mixing between left- and right-moving degrees of freedom in the
states that are in the cohomology of $\qtp$ than is possible 
in the on-shell cohomology of $Q+\bar Q$ where $Q$ and $\bar Q$ 
separately square to zero.  

We now describe a construction of $Q_T^{\prime}$ 
cohomology classes, which 
illustrates this non-trivial left-right mixing.
As given here,  the discussion will apply directly to the 
bosonic string case, and the NS-NS sector of the superstring. 
We expect that small 
 modifications will work for the R-R and R-NS sectors. 
The input will be  a pair of states, $A_{(-n)}$  of left moving 
ghost number $-n$  and $\bar B_{(-n)}$  of right moving ghost number $-n$, 
which 
satisfy a set of conditions: 
\eqn\contwo{\eqalign{  & {(\qp)}^{2n+1} A_{(-n)} = 0, \cr
            & (\bar Q^{\prime} )^{2n+1} \bB_{(-n)} = 0, \cr
   & \langle (A_{(-n)})^{\dagger} {(\qp)}^{2n} A_{(-n)} \rangle \ne 0, \cr
   & \langle (\bB_{(-n)})^{\dagger} {(\qpb)}^{2n} \bB_{(-n)} \rangle \ne 0. \cr
                   }}
In the actual cases of the bosonic string or the   NS-NS
sector, where we will apply this construction, the last two conditions 
in \contwo\ are not really necessary since 
any $Q_T^{\prime}$ closed state of ghost number zero is in cohomology, 
but we present the discussion in a way that does not use the $sl(2)$  
structure of section 2.6  and may thus be more general.  
                                    
Let us define 
\eqn\vardef{\eqalign{ &  A_{(-n+k)} \equiv {\qp}^{k} A_{(-n)},\cr
               & \bB_{(-n+k)} \equiv 
            ( \bar Q^{\prime} )^{k} \bB_{(-n)}. \cr }}
The subscripts denote ghost numbers. In some of our examples $A$ and 
$\bar B$ have spacetime indices which we have suppressed here. 

We will prove that the following state represents 
a non-trivial cohomology class: 
\eqn\state{ |s\rangle = 
\sum_{i=-n}^{n}  (-1)^{\sigma (i)}  A_{(i) }\bB_{(-i)}, }   
where $(-1)^{\sigma (i)} = \pm 1$ 
as determined by the following rule.
If $A_{(i)} $ is fermionic, then $\sigma(i+1) = \sigma (i) + 1 $. 
If $A_{(i)} $ is bosonic, then $\sigma (i+1) = \sigma (i)$. If we fix 
the sign of the first term to be positive say, this rule determines all 
the remaining signs.  
For $|s\rangle$ to be $\qp_T$-closed we need, for all $i$, 
\eqn\canc{  \qp  (  (-)^{\sigma(i)}  A_{(i)} \bar B_{(-i)} ) 
                      = -  \qpb ( (-1)^{\sigma (i+1)} A_{(i+1)} 
                                   \bB_{(-i+1)} \bigr)  }
It is easily checked that the definition of $\sigma(i)$ given above
guarantees this equality. Together with the first two equations in 
\contwo, this ensures that $ |s\rangle$ is $\qp_T$-closed. 
The norm of the state is given by 
\eqn\norm{\eqalign{  
 \langle s | s \rangle &=  (-1)^{\sigma (A_{(-n)}) + \sigma (n) + 1 } 
  \langle A_{(-n)}^{\dagger} Q^{2n} A_{(-n)} \rangle
\langle \bB_{(-n)}^{\dagger} (Q^{\prime})^{2n} \bB_{(-n)} \rangle \cr 
                       &= (-1)^{\sigma (A_0)+1} 
                  \langle A_0^{\dagger}A_0\rangle 
                   \langle \bar B_0^{\dagger}\bar B_0\rangle \cr
}} 
where $(-1)^{\sigma (A_{(-n)} )}$ is $+1$ or $-1$ 
depending on whether $A_{(-n)} $ is bosonic 
or fermionic, and $(-1)^{\sigma (n)}$ is  $+1$ or $-1$ depending on whether 
$n$ is even or odd. It is  interesting that the norm is determined 
only by the statistics and the norm of the ghost number zero states $A_{0}$
and $B_{0}$. If the state were $Q_T^{\prime}$-exact, it would be null, and  
by \contwo\ it follows that the state we have constructed is not 
$Q_T^{\prime}$ trivial.

We now outline the proof of \norm. 
The norm is : 
\eqn\calnorm{\eqalign{ & \langle s | s \rangle =  (-1)^{\sigma(0)}
         \langle (A_0\bB_0)^{\dagger} (A_0\bB_0) \rangle \cr
         & +  \sum_{i=1}^{n}  (-1)^{\sigma(i) + \sigma(-i)}
         \bigl( \langle (A_{(i)} \bB_{(-i)} )^{\dagger}  
         ( A_{(-i)} \bB_{(i)} ) \rangle    + 
   \langle ( A_{(-i)}B_{(i)} )^{\dagger}
           (A_{(i)}B_{(-i)}) \rangle \bigr)\cr 
         &= \langle A_0^{\dagger} A_0 \rangle \langle \bB_0^{\dagger}
                                                           \bB_0 \rangle \cr
         & + \sum_{i=1}^{n}  (-1)^{\sigma(i) + \sigma(-i)}
         \bigl( \langle A_{(i)}^{\dagger} A^{}_{(-i)}\rangle    
                 \langle \bB_{(-i)}^{\dagger}\bB^{}_{(i)}  \rangle  
                 +  \langle A_{(-i)}^{\dagger} A^{}_{(i)}\rangle    
                 \langle \bB_{(i)}^{\dagger}\bB^{}_{(-i)}  \rangle . \cr
}}
All other terms give zero because of ghost number conservation.   Using 
the equation 
\eqn\qton{\eqalign{ &  \langle A_{(-i)}^{\dagger} A_{(i)}\rangle \cr
          & =  \langle A_{(i)}^{\dagger} A_{(-i)}\rangle         \cr
          & = \langle A_{(-n)}^{\dagger} (\qp)^{ (n-i) + (n+i)} A_{(-n)}
 \rangle \cr 
          & = \langle A_{(-n)}^{\dagger} (\qp)^{ 2n} A_{(-n)} \rangle , 
}}
and its antichiral analog, it is clear that $\langle s|s\rangle$ is 
proportional to $$\langle A_{(-n)}^{\dagger} (\qp)^{ 2n} A_{(-n)} \rangle 
    \langle \bB_{(-n)}^{\dagger} (\qpb)^{ 2n} \bB_{(-n)} \rangle. $$ 
The description of the signs given after \state\  can be used to check the
coefficient in \norm.

Note that in the $n=0$ case this construction is just the tensor
product of a pair of states which are in cohomology of $\qp$ and $\qpb$
respectively. In this case, the last two conditions in \contwo\ reduce to 
the statement that the states are not null: 
\eqn\notnul{\eqalign{ &  \langle A_0^{\dagger} A_0 \rangle \ne 0 \,,\cr
         & \langle \bB_0^{\dagger} \bB_0 \rangle \ne 0 \,, \cr }}
which implies that they are not $\qp$ exact. 
In the following we will consider some examples of the above construction. 
It would be interesting to determine whether this construction gives a 
complete basis in the $Q^{\prime}_T$ off-shell cohomology. 

\subsec{Examples at low-lying levels} 
We now consider some examples at zero momentum, first in the case of the 
bosonic string.  At $L_0-1 = 2$, the state $A_{\mu \nu} 
\alpha_{-1}^{\mu}\alpha_{-1}^{\nu} |gh\rangle$  
is $Q^{\prime}$ closed if $A_{\mu \nu}$ is traceless. 
Taking this together with an analogous $\qpb$ closed state,
we can apply the above construction with $n=0$.
 
Another example is given by solutions to ${\qp}^3 |s\rangle = 0$, such as    
\eqn\qpthree{  b_{-2} |gh\rangle \,, \quad
 b_{-1}\alpha_{-1}^{\mu} |gh \rangle \,.} 
On both states ${\qp}^{3}$ is clearly zero because 
there is no ghost number $2$ state at this mass  level. Using 
the fact that ${\qp}^{2}$ is proportional to $M$, it 
also follows immediately that \contwo\ is satisfied. 
If we take 
$ A_{-1}^{\mu} =   b_{-1}\alpha_{-1}^{\mu} |gh\rangle$  
on the left and on the right 
$\bB_{-1}^{\nu} =  \bar b_{-1} \bar \alpha_{-1}^{\nu} |gh\rangle$
we get a set of states: 
\eqn\symtens{\eqalign{ & |s\rangle^{\mu, \nu}  = 
\bigl(  A_{(-1)}^{\mu} \bB_{(1)}^{\nu} 
 -  A_{(0)}^{\mu} \bB_{(0)}^{\nu} 
   - A_{(1)}^{\mu} B_{(-1)}^{\nu}  \bigr)\cr 
& = \bigl (   2 b_{-1}\alpha_{-1}^{\mu} \bar c_{-1} \bar \alpha_{-1}^{\nu}
  - \alpha_{-2}^{\mu} \bar \alpha_{-2}^{\nu} 
  - 2 c_{-1} \alpha_{-1}^{\mu} \bar b_{-1} \bar \alpha_{-1}^{\nu} 
\bigr )|gh \rangle \cr
& =  \bigl( -\alpha_{-2}^{\mu} \bar \alpha_{-2}^{\nu} - 
 2( c_{-1}\bar b_{-1} +  \bar c_{-1}b_{-1} )
\alpha_{-1}^{\mu} \alpha_{-1}^{\nu}\bigr ) |gh \rangle.   \cr
 }} 
Applying the same construction,  using on the left          
the state $A_{(-1)}^{\mu} =   b_{(-1)} \alpha_{-1}^{\mu} |gh\rangle$ 
and on the right $B_{(-1)} = b_{-2}|gh\rangle$ yields the off-shell 
physical state: 
\eqn\extwo{ \bigl[ \alpha_{-2}^{\mu} ( \bar \alpha_{-1}^{\nu} \bar\alpha_{-1\nu}
+ 6  \bar c_{-1} \bar b_{-1} ) + 4 \alpha_{-1}^{\mu} ( c_{-1}\bar b_{-2} + 2 
\bar c_{-2} b_{-1} )  \bigr] |gh\rangle.   } 

Let us also consider examples of this
construction in the NS-NS sector of the superstring at 
$L_0-1/2 = \bar L_0 -1/2 = 1/2$. (The first example is shown for its 
simplicity rather than its physical interest since it is projected out by 
GSO).  The conditions  \contwo\ are satisfied by the state 
$\beta_{-1/2} \psi_{-1/2}^{\mu} |gh;-1\rangle$.  Taking 
$A_{(-1)}^{\mu} = \beta_{-1/2} \psi_{-1/2}^{\mu}|gh;-1\rangle $
and $\bB_{(-1)}^{\nu} =\bar\beta_{-1/2}\bar\psi_{-1/2}^{\nu}|gh;-1\rangle $
we obtain the following state: 
\eqn\folsta{\eqalign{  |s \rangle^{\mu \nu} &=
 A_{(-1)}^{\mu}\bar B_{(1)}^{\nu}  + A_{(0)}^{\mu}\bar B_{(0) }^{\nu} -
 A^{\mu}_{(1)}\bar B_{(-1)}^{\nu} \cr 
&=  \bigl[  \alpha_{-1}^{\mu}\bar\alpha_{-1}^{\nu} 
   + 2 ( \beta_{-1/2} \bar \gamma_{-1/2} - \bar \beta_{-1/2} \gamma_{-1/2} ) 
 \psi_{-1/2}^{\mu}\bar \psi_{-1/2}^{\nu} \bigr]   |gh;-1, -1\rangle \,.\cr }}

One can also construct examples which survive the GSO projection. 
At $L_0-1/2= \bar L_0-1/2= 1$, take  $A_{(-1)}^{\mu} = 
( b_{-1} \psi_{-1/2}^{\mu}  
+ \beta_{-1/2} \alpha_{-1}^{\mu}  )  |gh;-1 \rangle$               
and $\bB_{(-1)} = \bar \beta_{-3/2} |gh;-1 \rangle $, to obtain
\eqn\nsnsex{\eqalign{ | s \rangle^{\mu} =
& A_{(-1)}^{\mu}\bar B_{(1)}^{}  + A_{(0)}^{\mu}\bar B_{(0) }^{} -
 A^{\mu}_{(1)}\bar B_{(-1)} \cr
          =&~ \bigl[   ( b_{-1} \psi_{-1/2}^{\mu}  + 
                     \beta_{-1/2} \alpha_{-1}^{\mu}  ) \bar  \gamma_{-3/2} \cr
              &~ + \psi_{-3/2}^{\mu} \bigl( \bar \alpha_{-1}\bar \psi_{-1/2} - 
             2  (\bar b_{-1} \bar \gamma_{-1/2} - 
                   \bar c_{-1/2} \bar \beta_{-1/2} ) \bigr )  \cr             
& ~ - ( c_{-1} \psi_{-1/2}^{\mu} +
               \gamma_{-1/2} \alpha_{-1}^{\mu}  ) \bar \beta_{-3/2} 
\bigr ] |gh;-1, -1\rangle \,. \cr }} 

\subsec{ Examples at arbitrary $n$ } 
Finally we give an example to show that solutions to \contwo\ exist for 
arbitrary $n$. Take the bosonic string, at zero momentum 
and $L_0=n (n+1)/2$. The state 
$A_{(-n)}= b_{-1}b_{-2} \cdots b_{-n} |gh\rangle$ satisfies
\eqn\an{ (\qp)^{2n} A_{(-n)} = C_1 (M)^{n} A_{(-n)} 
           = C_2 c_{-1}c_{-2} \cdots c_{-n} |gh\rangle, }
where $C_1$ and $C_2$ are non-zero constants. 
Now $\qp$ acting on this 
should have ghost number $n+1$. Such a state has at least 
$L_0 = (n+1)(n+2)/2 $, which implies $(\qp)^{2n+1} A_{(-n)}= 0$.  
Using the form of $(\qp)^{2n}$ \contwo\ is easily seen 
to be satisfied. Similar arguments show that in the 
NS sector an example is given by $(\beta_{-1/2})^n |gh; -1 \rangle$. 

\newsec{ Boundary states and cohomology  at higher mass levels}
In this section, we compare $Q_T^{\prime}$ cohomology with states that 
can be obtained by projecting known boundary states, to zero momentum 
and higher mass levels.

A large class of off-shell physical states can be written in terms 
of projections of  boundary states associated with simple D-brane 
configurations. For example consider for $I\ne J \ne K \ne L$, the state 
\eqn\st{ 
( \alpha_{-1}^I  \bar \alpha_{-1K} -  \alpha_{-1}^K  \bar \alpha_{-1I} )
 ( \alpha_{-1}^J  \bar \alpha_{-1L} -  \alpha_{-1}^L  \bar \alpha_{-1J} ) 
|gh\rangle , 
}
which can be obtained by applying the construction of the previous 
section at $n=0$.  It can also be expressed in terms of 
${\partial \over {\partial F_I^K}} {\partial \over {\partial F_J^L}}
|B (F_I^K,F_J^L ) \rangle $, where $|B (F_I^K,F_J^L ) \rangle$
is the boundary state of a D-brane with constant electromagnetic fields
$F_I^K$ and $F_J^L$. 

Another example is given by 
\eqn\sttwo{
 \bigl[  (\alpha_{-2}^I \bar\alpha_{-2J} - \alpha_{-2}^J \bar\alpha_{-2I} ) 
+ 2(\alpha_{-1}^I \bar\alpha_{-1J} - \alpha_{-1}^J \bar\alpha_{-1I} )  
 ( c_{-1}\bar b_{-1} +  \bar c_{-1} b_{-1} ) \bigr] |gh\rangle . }
States of this form can be obtained by construction of the previous section 
and can also be expressed in terms of  projections of 
${\partial \over {\partial F_I^J}} |B(F_I^J)\rangle$ with the derivative 
evaluated at zero field strength. 

There are also states, however, in $Q_T^{\prime}$ cohomology which cannot be 
obtained from known boundary states in this way.  An example is provided
by \extwo.  Analogously, in the NS-NS sector of the superstring there is the 
example \nsnsex . The non-trivial ghost structure of these states cannot be 
obtained from projections of standard boundary states, which 
satisfy the usual linear ghost boundary conditions:  
\eqn\ghstbd{\eqalign{  
&      c_{-n} = - \bar c_{n}, \cr
&      b_{-n} =  \bar b_{n}.  \cr }}

The existence of these exotic off-shell physical states has important 
implications for the physical interpretation of $\qp_T$
cohomology. It remains to be seen whether they correspond 
to a new class of boundary states. Another possibility 
is that these states have to be projected out of the off-shell 
cohomology in some way.

\newsec{Summary }
We have extended to the BRST context a definition of  
off-shell physical states, and described some of its basic 
properties and related vanishing theorems. We have 
discussed in some detail open and closed bosonic strings, and  type II  
strings in the NS-NS sector. Off-shell closed string physical states
exhibit a  mixing of left and right moving degrees of freedom that is 
not possible in the case of on-shell cohomology. 

At the massless level D-branes generate the entire zero-momentum cohomology 
of standard ghost number. This includes states which are continuations 
of the standard states of closed string perturbation theory around a flat 
background, as well as discrete states.
At zero momentum, states fall naturally in multiplets of 
$SO(9,1)$, whereas at non-zero momentum they fall in multiplets of 
$SO(8)$ in the massless case and $SO(9)$ in the massive case. 
This association of perturbative and discrete states into 
multiplets of a larger Lorentz group is 
reminiscent of the considerations
of \bars\ and it would be interesting to see if 
the $SO(9,1)$ structure can be simply 
interpreted in the framework of the higher dimensional formulations 
of string theory \refs{\town, \witM, \hull, \vaf, \kumat}. 

We have shown that a large class of non-trivial cohomology classes in 
this off-shell cohomology come from boundary states associated with 
D-branes.  At higher mass levels we also found  cohomology classes which 
do not come from standard boundary states.  If $Q+\bar Q$ closure and 
$\qp_T$ cohomology, are the only criteria selecting boundary states,  
this means that more general boundary states exist with a non-trivial 
mixing between matter and ghosts. An alternative possibility is that 
further conditions have to be imposed on boundary states. If the first 
possibility turns out to be correct it would be interesting to look for a  
group theoretic structure relating the new boundary states 
to previously known ones, generalizing the way $SO(9,1)$ relates 
known boundary states. If, on the other hand, the second possibility 
proves true, it remains to find the appropriate algebraic characterization 
of acceptable boundary states beyond $Q_T$ closure and $\qp_T$ cohomology. 

\bigbreak\bigskip\bigskip\centerline{{\bf Acknowledgements}}\nobreak
This work was supported in part by  DOE grant DE-FG02-91ER40671. 
We  thank J. Barbon, O. Ganor, A. Hashimoto, P. Horava for discussions.
 
\appendix{A}{Definition of off-shell cohomology for superstrings}
In the NS sector there are no superghost zero modes so the 
discussion is very similar  to the bosonic case. 
We will first discuss the open superstring.  We separate out the ghost zero 
modes: 
\eqn\gh{ Q = c_0 ( L_0-1/2 ) + b_0 M + Q^{\prime} } 
Using $Q^2 = 0 $, we find $ (Q^{\prime})^2 = - M(L_0-1/2)$.  
For off-shell states we can define $Q^{\prime}$ cohomology 
on the subspace where $M=0$. It contains analytic continuations of the 
DDF states.

Vanishing theorems for the NS sector can be proved by following the 
steps of section 2.6. The relevant $sl(2)$ algebra (see for example 
\lecdis\ and references there) is given by: 
\eqn\sltwons{\eqalign{
& M = -2 \sum_{n>0} nc_{-n}c_{n} + \gamma_{-n}\gamma_{n}, \cr 
& N = \sum_{n>0} -{1\over {2n}} b_{-n}b_{n} + 
                   {1\over 2} \beta_{-n} \beta_{n}, \cr 
& N_{gh} = \sum_{n\ne 0} : c_{-n}b_{n} : 
              -  : \beta_{-n}\gamma_{n} :. \cr }}  
As in section 2.6, we can show that any state 
of ghost number zero annihilated by $\qp$ is in cohomology. 
The extension of these statements to  the NS-NS sector 
of closed superstrings proceeds along the same lines. 

In the R sector of the open superstring, a simple definition of off-shell 
cohomology is again possible.  As before we make the ghost 
and superghost zero mode dependence explicit: 
\eqn\sep{ Q = c_0 L_0 + b_0 M -  \gamma_0^2b_0 + 
 \beta_0 G_0 + \gamma_0 K  + Q^{\prime}. } 
The nilpotence of $Q$ then implies that 
${Q^{\prime}}^2 =  -M L_0 + G_0 K$ (see \fof\ and references therein).
Away from the mass-shell condition we can consider 
$Q^{\prime}$ cohomology on the subspace
where it squares to zero. This $\qp$ operator 
commutes with $SO(9,1)$ so multiplets of this group 
can be expected at zero spacetime momentum. 

The BRST operator for the off-shell closed string case is defined 
by the sum of appropriate $Q^{\prime}$ and $\bar Q^{\prime}$
operators, and the subspace where the sum is zero is simply characterized 
in terms of $M$, $\bar M$, $K$ and $\bar K$.
Vanishing  theorems for off-shell cohomology  in the 
R sector of open strings, and the R-R and R-NS sectors 
of closed strings remain to be investigated.  

\listrefs
\end